
\input phyzzx


\def\eqalignalign#1{\vcenter{\openup1\jot
  \ialign{\strut\hfil
       $\displaystyle{##}$&$\displaystyle{{}##}$\quad&\hfil
       $\displaystyle{##}$&$\displaystyle{{}##}$\hfil\crcr
       #1\crcr}}}


\def\mPhi{m_{\Phi}}
\def\mgravitino{m_{3/2}}
\def\Phimin{\Phi_{\rm min}}
\def\msun{M_{\odot}}

\def\ls#1{_{\lower1.5pt\hbox{$\scriptstyle #1$}}}

\frontpagetrue


\let\picnaturalsize=N
\def\picsize{1.0in}
\def\picfilename{scipp_tree.eps}

\let\nopictures=Y

\ifx\nopictures Y\else{\ifx\epsfloaded Y\else\input epsf \fi
\let\epsfloaded=Y
{\line{\hbox{\ifx\picnaturalsize N\epsfxsize \picsize\fi
{\epsfbox{\picfilename}}}\hfill\vbox{


\hbox{MIT-CTP-2331}
\hbox{NSF-ITP-94-70}
\hbox{SCIPP 94-16}
\vskip1.2in
}
}}}\fi


\def\SCIPP{\centerline {\it Santa Cruz Institute for Particle Physics}
  \centerline{\it University of California, Santa Cruz, CA 95064}}

\def\MIT{\centerline {\it Laboratory for Nuclear Science and Department
         of Physics  }
         \centerline {\it Massachusetts Institute of Technology}
         \centerline {\it Cambridge, MA  02139} }

\def\ITP{\centerline {\it The Institute for Theoretical Physics}
         \centerline {\it Santa Barbara, CA 93106} }

\overfullrule 0pt

\pubtype{ T}     


\hfill\hbox{MIT-CTP-2331}

\hfill\hbox{NSF-ITP-94-70}

\hfill\hbox{SCIPP 94-16}


\title{Solving the Cosmological Moduli Problem with
Weak Scale Inflation}
\author{{Lisa Randall}
\foot{NSF Young Investigator Award,
Alfred P.~Sloan
Foundation Fellowship, DOE Outstanding Junior
Investigator Award. Supported in part
by DOE contract DE-AC02-76ER03069, by NSF grant PHY89-04035, and by
the Texas National Research Laboratory Commission
under grant RGFY92C6.}}
\MIT
\centerline{and}
\ITP
\vskip.6cm
\author{{Scott Thomas}
\foot{Supported in part by the U.S. Department of Energy
and the Texas National Research Laboratory Commission under grant
numbers RGFY 93-263 and RGFY 93-330.}}
\SCIPP
\vskip1cm
\vbox{
\centerline{\bf Abstract}

\parskip 0pt
\parindent 25pt
\overfullrule=0pt
\baselineskip=19pt
\tolerance 3500
\endpage
\pagenumber=1
\singlespace
\bigskip

Many models of supersymmetry breaking involve particles with
weak scale mass and Planck mass  suppressed couplings.
Coherent production of such particles in the early universe
destroys the successful predictions of nucleosynthesis.
We show that this problem may be solved by a brief period of
weak scale inflation. Furthermore the inflaton potential for
such an inflation naturally arises from the same assumptions
which lead to the cosmological problem.  Successful baryogenesis
and preservation of density fluctuations for large scale structure
formation are also possible in this scenario.
}

\vfill
\submit{Nuclear Physics B}
\vfill

\singlespace
\bigskip

\baselineskip=20pt

\chapter{Introduction}

An often discussed problem with  Planck scale physics is
the difficulty in deriving definite generic low energy
predictions.
Nonetheless, high energy scales are indirectly accessible
through their effects on cosmological evolution.
Although there are many uncertainties in this
evolution, at least two constraints
seem fairly reliable.
First, the remaining mass density in
stable particles should not ``overclose'' the universe
given the bounds on its age.
Second, the cosmology should allow
for a successful theory of nucleosynthesis.

\REF\bkn{T. Banks, D. Kaplan, and A. Nelson, Phys. Rev. D
{\bf 49}, 779 (1994).}

\REF\pol{G. Coughlan, W. Fischler, E. Kolb, S. Raby, and G. Ross,
Phys. Lett. B {\bf 131}, 59 (1983);
J. Ellis, D. V. Nanopoulos, and M. Quir\'os, Phys. Lett. B
{\bf 174}, 176 (1986);
R. de Carlos, J. A. Casas, F. Quevedo and E. Roulet, Phys. Lett. B
{\bf 318}, 447 (1993).}

In a recent paper, Banks, Kaplan, and Nelson\refmark{\bkn},
in a reincarnation of the well known Polonyi problem \refmark{\pol},
have shown the relevance of the second of these constraints
to models of dynamical supersymmetry
breaking, including those motivated
by superstring theory.
Many string models include directions in field space
which are flat in the supersymmetric limit and
couple to light fields only through Planck scale suppressed interactions.
If the potential for these flat directions is stabilized
by the same physics responsible for supersymmetry breaking,
a particle with mass of order a TeV and dangerously long
lifetime results.
Possible examples of such particles are the dilaton of string
theory, the massless gauge singlets of string compactifications, a Planck
scale coupled singlet responsible for supersymmetry breaking,
or a singlet field responsible
 for communicating supersymmetry breaking to the visible
sector. In this paper,
we refer to these fields collectively as moduli.
If one of the moduli starts with a Planck scale expectation
value, coherent oscillations about the minimum of the
potential dominate the energy density of the universe at
an early epoch.
Under naturalness assumptions for the couplings, one
finds that after decay of the moduli, the temperature
of the universe is too low to allow standard nucleosynthesis.
The conclusion is that the mass or couplings of
these fields is not consistent with naturalness assumptions
or that they are excluded\refmark{\bkn,\pol}.

In this paper, we investigate how this constraint
may be avoided.
We will show that the moduli problem
can be solved by a period of inflation with
Hubble constant of order of the weak scale.
Furthermore, such a period of inflation  is to be expected under
similar assumptions to those which give rise
to the cosmological problem because one
of the moduli or a flat direction in the
visible sector can act as the inflaton.
The resulting Hubble constant during inflation is dictated
by supersymmetry breaking, and is naturally of order the
weak scale.
Additional mass scales are therefore not introduced into
the theory. Furthermore, the same assumptions
which yield the dangerous moduli yield potentials
of the form required for either a ``new'' or a ``chaotic''
inflationary potential.
However, unlike the dangerous moduli
fields, the field driving inflation
is not necessarily weakly coupled.
The reheat temperature after inflation can then be much larger
than that for the dangerous moduli.
In fact, very natural choices for the inflaton parameters
reduce
the moduli amplitude sufficiently and generate
an acceptable reheat temperature.
We interpret our result as
indicating that cosmological constraints cannot be used to
eliminate models containing these Planck scale coupled moduli.

The structure of the paper is as follows. We first review the
cosmological problem of Planck scale coupled singlet fields connected
with the supersymmetry breaking sector, as discussed in
Ref. \refmark{\bkn}.
We discuss the various potential cosmological problems associated
with these TeV scale late decaying particles. We then discuss
potential ways to avoid these problems. Our primary focus will
be inflation, and we discuss the necessary properties of an inflationary
potential which can resolve the cosmological problems. We conclude
that a period of relatively late inflation, with the Hubble expansion
parameter of order the weak scale, readily solves the problem.
In Section 4,  we discuss the naturalness of the required
inflationary scenario.
We show
that such a period of
inflation might be expected under the same assumptions which
imply the existence of dangerous moduli fields.
Finally, we show our inflationary
scenario could naturally preserve density fluctuations
necessary to generate the structure of the universe if the duration
of inflation is sufficiently short.

\chapter{The Problem}
Supersymmetry provides a natural solution to the
large hierarchy between the weak and GUT or Planck scales.
In hidden sector models supersymmetry breaking
is transmitted to the low energy
visible sector through Planck scale suppressed interactions.
In nonrenormalizable hidden sector models \refmark{\bkn},
supersymmetry breaking also
vanishes when $M_p \to \infty$.
In exact supersymmetric theories ``flat'' directions in field
space for which the potential vanishes are common.
If the potential for these ``flat'' directions
is generated through the physics
associated with supersymmetry breaking, the potential
takes the form
$$
{\cal V}(\Phi) = \mgravitino^2 M^2 {\cal G}(|\Phi| / M)
\eqn\modulipot
$$
where $M \equiv M_p / \sqrt{8 \pi}$ is the reduced Planck mass,
$\mgravitino$ is the gravitino mass,
$\Phi$ represents a flat direction and ${\cal G}$ is some function.
The potential for the dangerous directions vanishes
in the flat space limit, $M \to \infty$,
since $\mgravitino \to 0$ in this limit.
Excitations about the minimum of the potential, $\Phimin$,
have mass
$\mPhi \sim \mgravitino$.
Notice the essential assumptions are that the potential vanishes
in the flat space limit and is proportional to $\mgravitino^2$.
In reality, it is difficult to construct models
for supersymmetry breaking of this sort.
However, we will show is that the assumed form of the potential
not only leads  to cosmological problems for moduli fields,
but can also yield an inflaton potential which naturally
solves the problem.

\REF\Dinecomment{String compactification moduli act as Higgs
particles near points of enhanced symmetry in the moduli space
where additional states become light.
If the minimum of the supersymmetry breaking potential coincides
with a point a enhanced symmetry, the moduli excitation
has Yukawa couplings to the additional light states.
The cosmological problems associated with the Planck scale
couplings would then be avoided for this type of moduli.
We thank M. Dine for this comment.}

Although the potential for flat directions vanishes for
$M \to \infty$,
due to the nonrenormalization theorems of supersymmetry,
this is not necessarily true of the couplings.
This means that
the couplings of flat directions to the low energy
visible sector can be large, even though all terms in the
potential \modulipot~are suppressed by powers of $M$.
For example, $D$-flat directions in the squark fields have Yukawa
couplings to other light fields.
However, in hidden sector models it is possible for
flat directions which are singlets under the low
energy gauge group
couple to other fields only through Planck mass suppressed interactions.
Examples of such fields include the dilaton and compactification
moduli\refmark{\Dinecomment} of string theory, and
singlet fields responsible for
breaking supersymmetry or communicating supersymmetry
breaking \refmark{\bkn}. These are the potentially dangerous
moduli fields.

There are several types of Planck mass suppressed couplings which might be
present. The moduli can couple either to gauge fields,
chiral fields of a hidden sector, or chiral fields of the visible
sector. By assumption, the coupling in each case is suppressed
by the Planck mass, but may be further suppressed.
For example, the coupling to gauge bosons might arise only at one loop,
due to an anomaly.
Decay to fermions through a current interaction
is suppressed by the fermion mass.
Decay through an explicit
helicity changing operator will have similar suppression.
The decay to scalars can be large but turns out to be suppressed
by numerical factors. We conclude that an estimate
for the decay rate of the moduli is at most
%
$$
\Gamma \sim {m_{\Phi}^3 \over 8 \pi M^2},
\eqn\rategg
$$
although the rate might be even smaller.
The slow decay rate for the moduli is the source of the
cosmological problem.

As discussed below,
relic moduli produced in the very early universe survive to
a dangerously late epoch.
These relic moduli could have arisen from  two distinct sources.
First, coherent
oscillations can give rise to a Bose condensate for the scalar component.
When the Hubble constant, $H$, reaches a value
$3H \sim \mPhi$,
the scalar component begins to oscillate coherently about
the minimum of the potential. 
The energy stored in this oscillation subsequently redshifts
like matter and can eventually dominate the energy density.
If the universe is radiation dominated when the oscillations
begin, the ratio of moduli number density in the
condensate, $n_{\Phi}= {1 \over 2} \mPhi \Phi^2$,
to entropy density,
$s=(2 \pi^2/45) g_{*s}T^3$,
where $T$ is the temperature when $3H \sim \mPhi$,
is
$$
{n_{\Phi} \over s} \sim 10^{-2}
     {\Phi_0^2 \over \sqrt{\mPhi M^3}}
\eqn\nscoh
$$
where $\Phi_0$ is the initial expectation value of $\Phi$
relative to $\Phimin$ and
$g_* \sim 275$ for the supersymmetric standard model.
Without any fine tuning,
the natural scale for $\Phi_0$ could reasonably be expected
to be of order $M$ for Planck scale coupled moduli,
which results in $n_{\Phi} / s \sim 10^6$.
If there are no entropy releases after the oscillations
begin, the moduli dominate the energy density at a temperature
$$
T \sim \mPhi {n_{\Phi} \over s}
\eqn\temprad
$$
If $\Phi_0 \sim M$, the coherent moduli dominate essentially
at the onset of oscillations.

\REF\gravprob{J. Ellis, A. Linde, and D. Nanopoulos,
Phys. Lett. B {\bf 118}, 59 (1982); D. Nanopoulos,
K. Olive, and M. Srednicki, Phys. Lett. B {\bf 127} 30 (1983);
J. Ellis, J. Kim, and D. Nanopoulos, Phys. Lett. B {\bf 145}, 181
(1984).}

The second source for both scalar and fermionic relic moduli is
thermal scattering in the plasma.
For $T \sim M$ the moduli are
in equilibrium with
a thermal number density, $n_{\Phi}/s \sim 1/ g_{*}$.
A significant entropy release, such as from
a standard inflationary scenario, would dilute this number density.
Even so, moduli can be produced
after the entropy release by thermal rescatterings.
Neglecting any initial number density, the Boltzmann
equation for the number density of incoherent moduli is
$$
\dot{n}_{\Phi} + 3 H n_{\Phi} = \sum_{i \geq j}
  \langle \sigma v \rangle_{ij}~ n_i n_j
\eqn\boltzmann
$$
where $\langle \sigma v \rangle_{ij}$ is the thermally averaged cross
section for
the initial states $i$ and $j$.
Typically, the two to two scattering
cross section dominates and is of order
$\sigma \sim \alpha / M^2$ where $\alpha$ is a gauge
coupling.
Integrating \boltzmann~from an initial temperature $T_I$
(an inflaton reheat temperature for example)
with this cross section gives an incoherent relic density of
$$
{ n_{\Phi} \over s} \sim  \alpha {T_I \over M}
\eqn\nsthermal
$$
For $T_I << M$ this is much smaller than a thermal number
density, and potentially quite insignificant
compared with the coherent number density \nscoh.
In Section 4, we show that
a weak scale inflation can
reduce the initial number density of both
coherent and incoherent moduli.
Thermal rescattering production of incoherent moduli after this
inflation
would
then be the dominant source.
However, so long as the reheat temperature after
inflation is sufficiently low to evade the similar problem
for gravitinos \refmark{\gravprob},
the incoherent production of moduli
will also be sufficiently suppressed.

\REF\scherrer{E. Kolb and R. Scherrer, Phys. Rev. D {\bf 25},
1481, (1982); R. Scherrer and M. Turner, Phys. Rev. D {\bf 331},
19 (1988); R. Scherrer and M. Turner, Phys. Rev. D {\bf 331},
33 (1988).}

\REF\ellisbound{D. Lindley, Ap. J. {\bf 294}, 1 (1985);
J. Ellis, D. Nanopoulos, and S. Sarkar,
Nucl. Phys. B {\bf 259}, 175 (1985);
S. Dimopolous, R. Esmailzadeh, L. Hall, and G. Starkman,
Nucl. Phys. B {\bf 311}, 699 (1988);
J. Ellis, G. Gelmini, J. Lopez, D. Nanopoulos, and S. Sarkar,
Nucl. Phys. B {\bf 373}, 399 (1992).}

\REF\reno{M. Reno and D. Seckel, Phys. Rev. D, {\bf 37}, 3441
(1988); G. Laxarides, R. Schaefer, D. Seckel, and Q. Shafi,
Nucl. Phys. B {\bf 346}, 193 (1990).}

A large relic density of moduli fields, independent
of the source, can be dangerous. If they dominate
the energy density before decaying, the universe
enters
a (moduli) matter
dominated era.
After the moduli
decay, the thermalized decay products
return the universe to a radiation dominated era.
The decay rate sets the scale for the ``reheat'' temperature
of the decay products.
In the sudden decay approximation the reheat temperature, $T_R$,
is defined by equating the lifetime, $\Gamma^{-1}$, with
the expansion time, $t= {2 \over 3} H^{-1}$,
where $H$ is the Hubble
constant at the time of decay,
$H= \pi \sqrt{g_{*} / 90}  T_R^2 / M$,
$$
T_R \sim \left( {40 \over  g_{*} \pi^2} \right)^{1/4}
\left( \Gamma M \right) ^{1/2}
\eqn\TReq
$$
Notice that $T_R$ is independent of
the initial moduli number density so long as the
moduli dominate the energy density before they decay.
With the decay rate \rategg
$$
T_R \sim 5 \left( \mPhi / {\rm TeV} \right)^{3/2}~ {\rm keV}
\eqn\TRest
$$
Such a low reheat temperature
is inconsistent with successful nucleosynthesis.
First,
a matter dominated era preceding $^4$He synthesis would
increase the expansion rate relative to the usual
radiation dominated scenario, thereby decreasing the
the neutron to proton ratio, $n/p$, and subsequently the
$^4$He fraction \refmark{\scherrer,\ellisbound}.
Reducing the initial $n_b /s$, where $n_b$ is the baryon density,
can postpone $^4$He synthesis, allowing for an acceptable
$^4$He fraction, but $^3$He and D are then
overproduced \refmark{\scherrer,\ellisbound}.
The second important constraint arising for relic decays occuring in
the latter stages of nucleosynthesis is from
the photodissociation and
photoproduction of light elements by decay
products \refmark{\scherrer,\ellisbound,\reno}.
These problems might be avoided for
$T_R > 1 {\rm MeV}$ since the weak interactions are in
equilibrium and the usual $n/p$ ratio can be obtained.
Even this is not sufficient though since $n/p$ can be modified
for $T_R > 1$ MeV when hadronic decay channels are available
(which is inevitable in
the thermalization process).
In fact, $T_R \gsim 6$ MeV is required for $\mPhi \sim$ 1 TeV in order
to ensure the $^4$He abundance is not too large \refmark{\reno}.
Clearly the low reheat temperature \TRest~ is not compatible
with the requirements for a standard nucleosynthesis scenario.
This is the cosmological problem for Planck scale coupled
moduli.
It is important to note that this is a problem for
{\it every} moduli of this type in the theory.


In fact, the problem is even more severe, in that
even if the moduli never dominates, the relic abundance
is severely restricted.
This is because even
a particle with small number density
which decays during or after nucleosynthesis is completed
can destroy existing nuclei, or change their abundance.
If the lifetime $\tau \lsim 10^4$ $s$ hadronic decay modes can increase
$n/p$, thereby overproducing $^4$He and
D + $^3$He \refmark{\ellisbound,\reno}.
If the lifetime $\tau \gsim 10^4$ $s$ electromagnetic cascades in the decay
can photodissociate and photoproduce light elements \refmark{\ellisbound}.
These bounds are only relevant if a significant fraction
of the moduli decay to the visible sector.
It is worth
noting that this can be avoided if the
moduli never dominate the energy density and
decay primarily to hidden sector fields. This
might occur for example if the couplings to the
visible sector were suppressed by an extra power of $M$.
The
exact bound in the case the moduli
do decay to the visible sector is a function of the mass of
the particle and
its lifetime, and varies considerably over the
parameter space.
The most stringent bound arises for $\tau \gsim 10^6$ $s$.
In this case for $\mPhi \sim 1$ TeV, $n_{\Phi}/s \lsim 10^{-15}$
is required in order not to overproduce
D + $^3$He \refmark{\ellisbound}.
This would require the initial
amplitude  of each  moduli field to be $\Phi_0 \lsim  10^{8}$ GeV
for nucleosynthesis to proceed as usual.

\REF\ewbaryo{For a review see A. Cohen, D. Kaplan, and
A. Nelson, Ann. Rev. Part. Nucl. Sci., {\bf 43}, 27 (1993).}

\REF\Affleck{I. Affleck and M. Dine, Nucl. Phys. B
{\bf 249}, 361 (1985).}

\REF\raby{The production of baryons from gravitino decay
in theories with broken R-parity
has been considered by J. Cline and S. Raby, Phys. Rev. D {\bf 43},
1781 (1991); and from saxino and axino decay by
S. Mollerach and E. Roulet Phys. Lett. B {\bf 281}, 303 (1992).}

\REF\lsp{The relic LSP abundance arising from gravitino decay
has been considered by L. Krauss, Nucl. Phys. B {\bf 227}, 556 (1983);
and from axino decay by K. Rajagopal, M. Turner, and F. Wilczek,
Nucl. Phys. B {\bf 358}, 447 (1991).}

We have so far considered the constraint on late decaying
particles from nucleosynthesis. There are other potential
problems, though these are more model dependent. Most
notable among these is the necessity for generating baryons.
  Baryon number
might be created by electroweak baryogenesis \refmark{\ewbaryo},
an Affleck-Dine mechanism \refmark{\Affleck},
or directly in the moduli decay \refmark{\raby}.
Generically, the first two
possibilities probably require a much higher reheat
temperature than that derived from nucleosynthesis. Because
sphaleron interactions are suppressed
below the electroweak scale, the reheat temperature is
required to be at least this order for the first possibility.
Although in principle, the Affleck-Dine mechanism can
work even with a much lower reheat temperature,
it is unlikely to meet all necessary constraints. In
particular, in theories with an unbroken R-parity,
if the reheat temperature is lower than the LSP mass by
about factor of ten, the
LSP will not be able to equilibrate.
If sufficiently
many baryons are created, under the most plausible
assumptions, the mass density of LSP's will be much
too high. Although models can be constructed
to avoid
this problem, the most likely scenario is that the
reheat temperature is high \refmark{\lsp}.

In what follows, we will assume the reheat temperature must
be of order 10 MeV, but will also consider the possibility
that it should be  higher. In fact, a high reheat temperature
can result very generally
 in the context of a natural inflationary scenario
discussed in section 4.

The cosmological problems associated with the low moduli
reheat temperature can be avoided if the naturalness assumptions
are relaxed.
The first possibility is
that the naive
estimate of the decay rate is too low.
For example,
the bound of $T_R \gsim$ 6 MeV can be reached with \rategg~for
a moduli mass of somewhat more than 100 TeV \refmark{\reno}.
This would happen if the moduli potential is stabilized
by something other than supersymmetry breaking physics,
or if the supersymmetry breaking scale is higher than
would be expected on the basis of naive estimates.
Alternatively, the rate could be large if the moduli
couplings to the visible sector are anomalously large.
A coupling a factor of $\gsim 10^3$ larger than that
assumed in \rategg~would result in an acceptable reheat temperature.
It is important to note that the mass and/or couplings of
{\it all} moduli must be large in order to solve the
problem in this way. This is certainly not what would
be expected on the basis of naive estimates.

The second possibility is that the
relic
moduli density  is lower
than the naive estimate.
If the moduli decay before dominating
the energy density of the universe, the requirement that
the universe be radiation dominated during nucleosynthesis
can be satisfied.
For $\mPhi \sim 1$ TeV and $\tau \gsim 10^3$ $s$,
the effect of the moduli energy
density on the expansion rate does not significantly
alter the yield of light elements if
$n_{\Phi} / s \lsim 10^{-7}$ \refmark{\scherrer}.
Including the effects of the  dilution of $n_b/s$, which would result
if the moduli decay predominantly to the visible sector,
increases this bound
to $n_{\Phi} / s \lsim 3 \times 10^{-7}
- 1 \times 10^{-8}$
for $\tau$ in the range $10^3 - 10^7$ $s$ \refmark{\ellisbound}.
This would require an initial amplitude of the oscillating
field to be $\Phi_0 \lsim 3 \times 10^{11}$ GeV.
However, in order
to avoid the more stringent bound
of $n_{\Phi} / s \lsim 10^{-15}$ from the D + $^3$He
abundance discussed above would require an initial amplitude
 $\Phi_0 \lsim 10^8$ GeV.
Because the precise bound is a complicated function of the
mass and lifetime, in this paper, we will assume that $n_{\Phi}/s$
must be less than $10^{-8}$, but
also consider the most stringent possible bound of
$10^{-15}$.

It is clearly extremely unnatural to assume
the initial relic densities of all moduli fields was so small.
In the next section, we show that a period of
weak scale inflation naturally reduces the relic density of all
moduli fields to an acceptable level. Unlike
the other possibilities we have mentioned, no
unnatural fine tuning of the moduli fields will be necessary.

\chapter{Diluting Moduli with Inflation}

The cosmological problems of the moduli
will be solved only if 1) any preexisting
thermal moduli are diluted, 2) the production
of moduli by thermal rescatterings is
small, and 3) the amplitudes
of all scalar moduli are sufficiently small
to eliminate excessive relic  coherent moduli.
A standard inflationary scenario solves the first
two problems; in this section we
 show that inflation can also solve
the last problem, although this is most naturally
achieved if the Hubble constant during inflation
is low, of order of the electroweak scale. In this
section, we demonstrate how inflation solves the
cosmological problems. In subsequent sections,
we will address the
questions of naturalness and density perturbations.

\REF\dimhall{S. Dimopoulos and L. H. Hall, Phys. Rev. Lett. {\bf 60},
1899 (1988).}

First consider the classical evolution of the scalar moduli.
During inflation the Hubble constant is approximately
time independent.
The classical equation of motion for the zero mode of a
moduli is then just that of a damped oscillator
$$
\ddot{\Phi} + 3 H \dot{\Phi} + {\cal V}^{\prime}(\Phi) =0
\eqn\dho
$$
For $\Phi \lsim M$ the harmonic part of the potential should be
most important.
The solutions to \dho~in the over-, critically-,
and under-damped cases are then
$$
\eqalignalign{
\Phi(N) & \simeq \Phi_0 e^{-(\mPhi^2 / 3 H^2) N} \hskip1in &
    (H >> \mPhi)
\cr
\Phi(N) & = \Phi_0 \left( 1 + {3 \over 2} N \right)
    e^{- {3 \over 2} N} & (H={2 \over 3} \mPhi)
\cr
\Phi(N) & \simeq \Phi_0 e^{ - {3 \over 2} N} & (H<< \mPhi)
\cr }
\eqn\amplitude
$$
where $N \equiv Ht$ is the number of $e$-foldings since the
beginning of inflation and $\Phi_0$ is the moduli
expectation value for $N=0$.
Depending on the number of $e$-foldings, $\Phi$ can be
significantly reduced.
In the overdamped case $(H>>\mPhi)$,
to first approximation $\Phi$ is
unchanged.
A standard inflation with $H \sim 10^{13}$ GeV and $N \sim 60$
required to solve the horizon and flatness problems therefore
does not significantly alter the moduli amplitude.
However, if $N >> (H/ \mPhi)^2$ the classical amplitude is reduced.
In the highly overdamped case this would require
an extremely fine-tuned inflationary
potential to obtain such a large number of
$e$-foldings.\foot{We note that this solution to the cosmological moduli
problem is similar to the solution of the cosmological
axion problem presented in Ref. \refmark{\dimhall}.
However, an enormous number of $e$-foldings is required when
the inflation scale is not extremely low.}
Measured by the number of required $e$-foldings,
the underdamped $(H << \mPhi)$
case is clearly much more efficient in reducing the amplitude.
In this case, the coherent moduli begin
oscillating
before the inflationary phase.
The reduction in  amplitude then amounts to a dilution of
the number density in the coherent condensate by
$e^{-3N}$, just as for any number density.

\REF\desitter{T. Bunch and P. Davies, Proc. R. Soc. A
{\bf 360}, 117 (1978);
A. Linde, Phys. Lett. B {\bf 116}, 335 (1982);
A. Starobinsky, Phys. Lett. B {\bf 117}, 175 (1982);
A. Linde, Phys. Lett. B {\bf 131}, 330 (1983).}

\REF\chaosnote{Even a
chaotic inflationary
scenario with initial vacuum energy of order
$M^4$ only would have sufficiently many $e$-foldings
to damp the moduli fields if $H \gsim 10^8$ GeV).}

In addition to the classical evolution, the moduli fields
undergo quantum deSitter fluctuations in the inflationary
phase.
For a harmonic potential the classical and quantum evolutions
decouple.
In the underdamped case the quantum fluctuations, $\delta \Phi$, are
exponentially suppressed and unimportant.
In the overdamped case initially the classical evolution is frozen and
the fluctuations  grow
as for a massless field,
$\sqrt{ \langle (\delta \Phi)^2 \rangle} \simeq \sqrt{N} H / 2 \pi$
\refmark{\desitter}.
For $N \gsim (H / \mPhi)^2$ the growth of fluctuations
ceases and the low
frequency modes have a thermal occupation with
$\langle (\delta \Phi)^2 \rangle \simeq 3 H^4 / 8 \pi^2 \mPhi^2$
\refmark{\desitter}.
The quantum fluctuations represent a lower bound on the
rms value of $\Phi$ at the end of inflation, thereby constraining
the overdamped scenario.
Assuming the moduli begins oscillating in the radiation era
following inflaton decay, $H \lsim 5 \times 10^7$ GeV is
required during inflation in order that the rms fluctuations
in $\Phi$ satisfy the bound of
$n_{\Phi} / s \lsim 10^{-8}$.
Although an inflation with
$\mPhi \lsim H \lsim 5 \times 10^7$ GeV
and $N >> (H / \mPhi)^2$ could in principle solve the moduli
problem,  there is no natural source for $H$ is this
range and a huge
number of $e$-foldings would be  required.\foot{However a chaotic
inflationary potential of the form
${\cal V} = (\mu / M)^2 \phi^4$ has a maximum number of $e$-foldings
$M/ \mu$ for ${\cal V} \sim M^4$.
The moduli fields will be damped most efficiently when $H$ has fallen
to $H \sim \mu$  at which point $N \gsim (\mu / \mPhi)^2$ is required.
This works only for $\mu \lsim 10^8$ Gev.}
We therefore disregard the overdamped
scenario and focus on the underdamped and critically damped
cases in what follows.

The inflationary phase ends when the inflaton
starts to undergo coherent oscillations.
The universe then enters
an inflaton matter dominated era.
When the Hubble parameter is of order of the inflaton
decay rate, $\Gamma_I$,
the inflaton decays, reheating the universe to a temperature
$T_r^4 \sim (\Gamma_I M)^{1/2}/ g_*^{1/4}$.
Clearly this temperature depends on the strength
of the inflaton coupling. The reheat temperature should
not be too large; otherwise there is a problem with
thermal rescatterings producing gravitinos, moduli, and
perhaps other relic particles.
In the underdamped scenario this bound is readily satisfied;
the maximum reheat temperature if the inflaton is strongly
coupled and decays immediately following inflation
is $T_{r,{\rm max}} \sim (H M)^{1/2}/ g_*^{1/4}
\lsim 10^{10}$ GeV (since $H \lsim \mPhi$).
The inflaton is unlikely to be strongly coupled so the
actual reheat temperature will be much less.
In fact the stronger bound on this scenario will be
achieving a sufficiently high reheat temperature
to avoid the the nucleosynthesis bounds discussed in the
previous section.
However we will see in the next section that there  exist
models with this property.

The relic moduli density after inflation can now be determined
in the underdamped scenario.
During inflation, the coherent number density is diluted by
a factor $e^{-3N}$.
After inflation the ratio of moduli to inflaton number density
is constant until the inflaton decays.
Since the inflaton decay is the source of entropy for the
subsequent radiation era, the ratio of moduli to entropy
density is
$$
{n_{\Phi} \over s} \sim
{\mPhi \Phi_0^2e^{-3N} \over 2 (2 \pi^2 / 45) g_*T_r^3}
   \left( { R_{osc}} \over R_{r} \right)^3 \sim
{\mPhi T_r \over 72 m_I^2} { \Phi_0^2 e^{-3N} \over M^2}
\eqn\nsinflation
$$
where $m_I$ is the inflaton mass (of order $H$ during inflation
for a natural potential), and $R_{osc}$ and $R_r$ are the
scale factors at the epoch when the inflaton begins oscillating
and decays respectively.
The result is essentially the same for the critically damped case.
The required number of $e$-foldings to sufficiently
dilute the moduli fields depends on the reheat temperature.
For the most natural models described in the next section this
lies between 10 MeV and $10^5$ GeV.
As an example, for the moderately underdamped case of
$m_I \sim H \sim {1 \over 3} \mPhi$,
and assuming the initial moduli amplitude before inflation
was $\sim M$,
the required number of $e$-foldings is 2-7 for $T_r \sim 10$ MeV
(for the bound on $n_{\Phi}/s$ between $10^{-8}$ and
$10^{-15}$) and
6-12 for $T_r \sim 10^{5}$ GeV.
These numbers are low,
as a significant dilution occurs during the inflaton reheat stage.

It is important to note that although the inflaton decay contributes
to the moduli dilution, it alone is not sufficient.
This is because for $\Phi_0 \sim M$, without the dilution factor
$e^{-3N}$, the moduli and inflaton energy  densities are roughly
equal after the moduli begin oscillating (irrespective of when
the oscillations begin).
After the inflaton decays the energy in the
thermalized decay products redshifts and
the moduli dominate the energy density below $T \sim T_r$.
Although this is a much lower temperature than that
from \nscoh~and \temprad, $T_r \gsim 6$ MeV is necessary
for the inflaton
in order
to obtain successful nucleosynthesis as discussed in section 2.
The requirement that the universe be radiation dominated
during nucleosynthesis could then not be satisfied.
A brief period of weak scale inflation
in which the moduli amplitudes are reduced
is therefore necessary to
solve the moduli problem.

\chapter{Inevitable Inflation}

\REF\chaotic{A. Linde, Phys. Lett. B {\bf 129}, 177 (1983).}

\REF\new{A. Linde, Phys. Lett. B {\bf 108}, 389 (1981);
A. Albrecht and P. Steinhardt, Phys. Rev. Lett. {\bf 48},
1220 (1982).}

\REF\conditions{P. Steinhardt and M. Turner, Phys. Rev. D {\bf 29},
2162 (1984).}

We next consider the question of the naturalness of the
type of inflationary scenario outlined above, namely
one with Hubble parameter of order the weak scale and
reheat temperature of at least 10 MeV. We will show
first of all that the assumed form of the potential
for a ``flat'' direction
is one that can readily be associated with ``chaotic''
inflation \refmark{\chaotic},
although with small fine tuning
it can be associated with a ``new'' inflationary
scenario \refmark{\new}.
Furthermore, the expected order of magnitude for the Hubble parameter
is $m_{3/2}$.
This is exactly what is required to
efficiently dilute the dangerous moduli fields.
The second requirement of our inflationary scenario
was a sufficiently high reheat temperature, which
should be at least 10 MeV, and probably higher (to account
for the generation of baryons).
A sufficiently high
reheat temperature is naturally obtained
if the inflaton has renormalizable
couplings to the visible sector.
We will consider examples
of such inflatons comprised of either standard model or nonstandard
model
supersymmetric flat directions.

In order for the Hubble constant during inflation to be
of order $\mgravitino$ the vacuum energy density, $V$,
must be of order $\mgravitino^2 M^2$. Now suppose
the inflaton corresponds to a particle with potential
 \modulipot.
The initial value of the flat direction is expected
to be  of order $M$, corresponding
to $V \sim \mgravitino^2 M^2$.
If inflation does occur the Hubble parameter,
$H \sim V^{1/2} / M$,
is then naturally
expected to of order $\mgravitino$,
and is precisely of the order of magnitude for which we expect
inflation to be most efficient in reducing the number
density of potentially dangerous moduli.
No new mass scales are introduced into the theory.
This
is the most compelling feature of weak scale inflation as a solution
to the cosmological moduli problem.

Now consider the question of the likelihood of inflation.
The existence of compelling inflationary models is not
yet satisfactorily resolved. The two favorite scenarios
are based on either a ``new'' inflationary potential \refmark{\new}
which requires the potential be sufficiently flat for slow
roll but with sufficient curvature for reasonable reheating,
or the ``chaotic'' inflationary scenario \refmark{\chaotic}
which requires
assumptions about initial conditions,
and a small parameter in order to generate sufficiently small
density fluctuations.
As a  measure of  inflation consider the
number of $e$-foldings which result
with the assumed form of our potential,
$$
{\cal V}=m_{3/2}^2 M^2 {\cal G}(\phi/M)
\eqn\inflatonpot
$$
where $\phi$ is now the inflaton.
Using the slow roll equation $3 H \dot{\phi} \simeq -{\cal V}'$
the number of $e$-foldings is
$$
N=\int H dt =
\int dt {\sqrt{{\cal V}/3} \over M}
=\int d\phi {{\cal V} \over M^2 {\cal V}'}\approx
{\Delta \phi  \over M}{{\cal G} \over {\cal G}'}
\eqn\nefoldings
$$
where $\Delta \phi$ is the total change in $\phi$ during inflation.
New inflation would obtain sufficiently many $e$-foldings through
a flat potential.
Specifically, the conditions for new inflation
with the potential $\inflatonpot$ are
${\cal G}'' << 3$
$ {\cal G}$
and ${\cal G}' << \sqrt{6} {\cal G}$ \refmark{\conditions}.
Although the
potential must be somewhat fine tuned in order to get
inflation, the solution of the cosmological moduli
problem only requires a small number of $e$-foldings,
so this might not be very unnatural. Furthermore,
we will see in the next section that the most favored
scenarios are indeed those with small numbers of $e$-foldings
since density fluctuations generated at an earlier epoch
necessary
for galaxy formation can then be preserved.


\REF\eternal{A. Linde, Phys. Lett. B {\bf 175}, 395 (1986).}

Alternatively, the initial value of the inflaton field might be
somewhat greater than $M$.
This leads to a ``chaotic'' inflationary scenario \refmark{\chaotic}.
In general, chaotic inflation requires a small parameter
so that  density fluctuations are sufficiently small.
The assumed form of the potential \inflatonpot~(that which generates
both the cosmological problem, and potentially its solution)
does contain a small parameter.
For example, an expansion of ${\cal G}$
would give a quartic coupling of order $(m_{3/2}/M)^2$.
Chaotic inflation makes no assumption about
${\cal G}''$, but
only requires that $({\cal G}'/{\cal G})^2 < 1$
which follows from $\dot{H} < H^2$.
For a potential with polynomial behavior at large
field this requires $\phi_0>M$, where $\phi_0$ is the field
value during inflation. In fact, if only a few $e$-foldings
are required, the polynomial form is not necessary.
We refer to weak scale inflation of this type as
``inevitable'' inflation since
nothing is assumed other than the form of the
potential which follows from the assumptions of a nonrenormalizable
hidden sector.
Notice that our scenario is distinguished from that of Linde
in that we do not assume the potential energy is as large as
$M^4$ during inflation \refmark{\chaotic},
but rather only $m_{3/2}^2 M^2$ corresponding to ${\cal G} \sim 1$.
It is hard to estimate the likelihood of one initial
condition over another.

If the potential takes
a polynomial form, it will allow the possibility of ``eternal''
inflation \refmark{\eternal}
so long as  the deSitter fluctuations, of amplitude
of order $H$, are  greater than the change in $\phi$ according
to its classical evolution in a Hubble time, $V'/H^2$, which
is equivalent to
${\cal G}'/{\cal G}^{3/2}<m_{3/2}/M$.
Unlike the condition for chaotic inflation,
$(( {\cal G} / {\cal G}')^2 < 1)$, this inequality  involves a small number,
$\mgravitino / M$.
It can therefore be satisfied by potentials which are large for
some field value (e.g. a polynomial potential).
So a reasonably flat potential allows
the possibility of chaotic inflation but not eternal inflation,
whereas a polynomial potential allows both.
As discussed in the next section
the preservation of density fluctuations requires a small
number
of $e$-foldings.
The chaotic inflation we envision therefore corresponds
to initial energy density $V \sim m_{3/2}^2 M^2$ but not much
bigger. If eternal inflation is permitted, this initial condition
might be unlikely; this question of initial conditions is
however beyond the scope of this paper.
Nonetheless,
we find it compelling that the assumed form of the potential
allows for ``inevitable" inflation, and automatically
contains the small parameter required for
consistent chaotic inflation. 


Having shown that it is quite natural to expect
a flat direction of the supersymmetric potential
to play the role of an inflaton with associated
Hubble parameter of order $m_{3/2}$, we now turn
to the question of the reheat temperature after
inflation.  Clearly, if the inflaton
were identified with one of the dangerous moduli,
the reheat temperature would be of order 5 keV
which is much too low for successful nucleosynthesis.
As with the moduli fields themselves, one can imagine that
either the mass or couplings are larger than
would be naively estimated on dimensional grounds.
This scenario has the advantage
that it only requires tuning the parameters
of a single field. However, a very massive inflaton would
probably mean that the Hubble parameter during inflation
is larger than the mass of the dangerous moduli fields,
which would also imply inefficient damping of the moduli amplitude.

A much better scenario would require the inflaton to have
renormalizable couplings (not suppressed by $M$) to the
visible sector.  This is consistent with the fact that
the inflaton is a flat direction in the supersymmetric limit
because of the supersymmetric
nonrenormalization theorems.
Gauge couplings would
arise for either a nonstandard model particle with color or electroweak
gauge couplings, or for a flat direction in the visible sector.
Whether the former type of flat direction actually
occurs depends on the underlying model.
One might worry about the finite temperature corrections to the
potential;
however, so long as the field value is large,
the relevant thermal degrees of freedom are
heavy and their thermal abundance is exponentially
suppressed.
We first show the possibility of
flat directions in the standard supersymmetric
sector. These can have both gauge and Yukawa couplings.

\REF\lance{L. Dixon, in {\it Superstrings, Unified Theories, and
Cosmology 1987}, Proceedings of the Summer Workshop in High energy
Physics and Cosmology, eds. G. Furlan {\it et. al.}, (World
Scientific, Singapore, 1987) 
.}

As discussed in Ref. \refmark{\Affleck}~there are many flat
directions which occur in the renormalizable
supersymmetric standard model
(these are flat with respect to both $F$ and $D$ terms in
the potential).  However, for our scenario to be viable,
the $F$ term must vanish along the flat direction to very high
order in $1/M$.
In fact, it is sufficient to show that there are no
Planck suppressed superpotential terms. This is
because all nonPlanck suppressed terms in the potential
are explicitly proportional to the  $F$ component
of the flat direction.
It might be that an underlying superstring model protects
against the generation of all possible Planck suppressed
holomorphic potential terms which are consistent with the
symmetries \refmark{\lance}.
It is also possible for field theory symmetries
to guarantee that a flat direction is
flat to arbitrary order in $M$ in the supersymmetric limit.
A simple example in terms of
the standard supersymmetric model fields
is $Q^1_1 = v, L_2^1 = v$, and $\bar{d}^2=v$,
with a  conserved  global $U(1)_{B-L}$.
Lower indices are for $SU(2)_L$, and upper indices are for generation.
It is
easy to see that any gauge invariant operator
with nonzero $F$ component
which can be constructed in the superpotential
does not respect the global symmetry.
The global symmetry may not be gauged in this example since
the direction would then not be $D$ flat. This appears
to be generally true; for ordinary (non $R$) symmetries,
the conditions for $D$ flatness allow for nonvanishing
nonrenormalizable terms in the superpotential, so that
these directions will not be $F$ flat.



If the inflaton field has either
gauge couplings or a  direct Yukawa coupling
to visible field particles,
a one loop diagram will generate a $D$ type term of
the form \refmark{\Affleck}
$$
{g^2  \over \langle \phi \rangle} \int d^4 \theta
     \chi^\dagger \chi \phi
\eqn\
$$
which includes component couplings of the form
$$
{g^2 \over \langle \phi \rangle} \phi \psi_\chi \partial \psi_\chi^\dagger\, ,
$$
where the fermion may or may not be the fermionic component of
the $\phi$ field, and
loop factors are absorbed in the definition
of the coupling $g$, which may be a gauge or Yukawa coupling.
The inflaton decays when the Hubble parameter,
$H \sim m_\phi \langle \phi \rangle/M$
is approximately equal to the width $\Gamma \sim g^4 m_\phi^3/ \langle
\phi \rangle ^2$.  This
yields a reheat temperature, $T_r \sim \sqrt{\Gamma M} \sim
g^{2/3} m_\phi^{5/6} M^{1/6} \sim g^{2/3} 10^5 {\rm GeV}$.
This might be as large as
an order of magnitude bigger than the weak scale.

When there is a Yukawa
coupling  $\lambda \phi \phi_1 \phi_2$
in the superpotential which
couples the inflaton $\phi$ directly to $\phi_1$ and $\phi_2$,
there would also  be a direct decay into the associated fields,
providing they are lighter than the inflaton. However, so long
as $\phi$ is large, $\phi_1$ and $\phi_2$ will be heavy and
the decay is kinematically forbidden.
Nonetheless,  the amplitude of $\phi$ decreases
as the number density decreases during the (inflaton) matter dominated
era following inflation. When $\langle \phi \rangle$
becomes sufficiently small to kinematically allow the decay,
we expect $\phi$ to decay essentially
instantaneously, so long as $\Gamma>H$
at this time. Assuming the Yukawa coupling is $\lambda$,
the decay rate would be of order $\Gamma \sim \lambda^2 m_\phi/16 \pi^2$.
The Hubble constant at this time is
of order $m_\phi \langle \phi \rangle/M$.
Since $\lambda \langle \phi \rangle< m_{\phi}$ in order
for the decay to be allowed, we find the Hubble constant
at the time of decay will be less than $m_{\phi}^2/\lambda M$. So long
as $\lambda $ is not too small  (greater than about $10^{-5}$
so that $\Gamma>H$ when kinematically allowed), the
decay will occur essentially
immediately. If this is the dominant $\phi$
decay mode, the reheat temperature would be expected to be
about $T_r \sim m_\phi/\sqrt{\lambda}$. This can be as
large as
several orders of magnitude larger than the weak scale.

We conclude that it is very natural to have inflation with
inflaton amplitude a few times $M$, a Hubble
constant of order the weak scale, and reheat temperature at
or above the weak scale. This does not require fine tuning
of parameters.
Multiple flat directions may contribute to  inflation.
The most significant requirement is that
the flat direction which inflates the longest is one with
renormalizable couplings to the visible sector, so that
the reheat temperature is sufficiently high. All other requirements
fall naturally from the assumption that inflation is
associated with a flat direction of the supersymmetric theory.

The only requirement of inflation which cannot be easily met
without fine tuning the potential is the generation of density
fluctuations. However, if the number of $e$-foldings is not
too large, late inflation will preserve density fluctuations
which were produced during an earlier cosmological epoch.
We discuss this in the following section.

\chapter{Retaining Density Fluctuations During Inevitable Inflation}

\REF\spectrum{A. Guth and S.-Y. Pi, Phys. Rev. Lett. {\bf 49},
1110 (1982); A. Starodindkii, Phys. Lett. B {\bf 117}, 175 (1982);
S. Hawking, Phys. Lett. B {\bf 115}, 295 (1982);
J. Bardeen, P. Steinhardt, and M. Turner, Phys. Rev. D {\bf 28},
679 (1983).}

\REF\joel{J. Primack, {\it private communication}.}

In the previous section, we showed that it is fairly natural to
expect an inflation with $H \sim  m_{3/2}$. Such a late
inflation solves the cosmological moduli problem, and
could in principle
solve the horizon and flatness problems with sufficiently
many $e$-foldings.
However, the observed density fluctuations
of $ \delta \rho / \rho \sim 10^{-5}$ could not be generated
without an extreme
fine tuning of the potential.
For such a small Hubble constant the density fluctuations are dominated
by the deSitter fluctuations of the inflaton,
for which \refmark\spectrum
$$
{\delta \rho \over \rho} \approx {H^2 \over \dot{\phi}}\approx
{H^3 \over {\cal V}'}
\eqn\deltarho
$$
where the last equality is from the slow roll condition.
Without any fine tuning of the potential,
$\delta\rho/\rho$ would be expected
to be of order $m_{3/2}/M$. This means
that the observed density fluctuations require a fine
tuning of ${\cal G}'/{\cal G}^{3/2}$ at the level of
one part in $10^{10}$.
This is clearly not natural.
A more reasonable assumption is
that density fluctuations were created prior to the period
of late inflation, presumably by an earlier inflationary
epoch with much larger Hubble parameter.
In this case,
all that is required of late inflation is that it does
not destroy the density fluctuations which are necessary to
produce the large scale structure of the universe.

All fluctuations which pass outside the horizon during the weak
scale inflation have amplitude given by \deltarho,
which in a natural scenario is quite small.
The requirement is therefore that all these fluctuations
come back inside the horizon before a time relevant to the
growth of large scale cosmological density fluctuations.
Any primordial fluctuations on scales larger than $H^{-1}$
at the beginning of weak scale inflation are unaffected.

The smallest objects
which could reasonably be related to cosmological
density fluctuations are dwarf galaxies and Lyman alpha clouds
with mass of order $10^6 \msun$ \refmark\joel.
Of course, the growth of density fluctuations
on this scale are not measured directly; the necessity for structure
formation from cosmological fluctuations
is obtained from assuming a scenario for galaxy formation.
It is possible this criterion is more conservative than required.

\REF\joelpaper{J. Primack, in {\it Proceedings of BCSPIN International
School on Particle Physics and Cosmology} ed. J. Pati, (World
Scientific, 1994).}

The largest scale affected by the weak scale inflation is that
which goes outside the horizon at the beginning of inflation,
with size $H^{-1}$.
During inflation it is stretched by a factor $e^N$.
After inflation it is further stretched by the expansion of
the universe in the inflaton matter dominated era and radiation
era following the inflaton decay.
The horizon also grows after inflation, eventually encompassing
this scale.
The requirement is  that the
size of the largest fluctuation affected
by late inflation
 must be smaller than the Hubble size when
$10^6 \msun$ of nonrelativitic matter enters the horizon,
$$
H_{10^6}^{-1}  \gsim H^{-1} e^N
   \left( {V \over g_* T_r^4}\right)^{1/3}
   \left({T_r \over T_{10^6}}\right)
\eqn\saverho
$$
where $V \sim \mgravitino^2 M^2$ is the vacuum energy at the end
of inflation, $H$ is the Hubble constant during inflation,
and $T_{10^6} \sim 5$ keV is the temperature when $10^6 \msun$
of nonrelativistic matter passes inside the horizon with associate
Hubble constant $H_{10^6} \sim g_*^{1/2} T_{10^6}^2 /M$ \refmark{\joelpaper}.
The factors in \saverho~are as follows.
The first
factor is the size of the initial Hubble patch stretched by inflation.
The second is the amount by which this patch is stretched during the
matter dominated era between the end of inflation and inflaton
decay.
The third is the stretch during the radiation era between inflaton
reheating and $T_{10^6}$.

The criterion \saverho~amounts to a bound on the number of $e$-foldings
for a given inflaton reheat temperature.
If the reheat temperature is as low as 10 MeV,
$N \lsim 25$ is required.
If $T_r \sim 10^5 {\rm GeV}$, a less
stringent bound of $N\lsim 30$
is obtained.
Notice that
with this bound on the number of $e$-foldings, the cosmological
moduli problem is solved.
However, with this small number
of $e$-foldings,  a prior period of inflation is
required to solve the horizon and flatness problems,
as well as to imprint density fluctuations on large scales.

\chapter{Conclusions}

The cosmological moduli problem presents a potentially serious
threat to hidden sector models of supersymmetry breaking with
Planck scale suppressed couplings and weak scale masses.
The low reheat temperature after moduli decay is not
sufficient to obtain a successful period of nucleosynthesis.
The bounds from nucleosynthesis might be avoided if naturalness
assumptions about all the moduli masses and/or couplings are
relaxed.
Even then, baryon production
could be a significant constraint.
In this paper we have show that a period of inflation
with small enough Hubble constant can reduce the scalar
moduli amplitude to acceptable levels.
The most efficient dilution occurs for a Hubble constant
of order or smaller than the weak scale.
The most important conclusion here is that a ``flat''
direction can naturally give rise to inflation with just
such a Hubble constant.
No new mass scales are introduced into the theory.
This is the most compelling aspect of weak
scale inflation.
The flatness of the potential which gives rise to the
problem could also lead  to the solution.

The form of the inflation may be associated with a
``new'' inflationary scenario with a moderate tuning
of the potential to achieve slow roll for Planck
scale field values.
Alternately a ``chaotic'' scenario can result if the initial
field in the flat direction is somewhat larger than the
Planck mass.
The supersymmetry breaking induced potential for the flat directions
therefore leads to an ``inevitable'' inflationary scenario
with weak scale Hubble constant.
One of the dangerous moduli can act as the inflaton for such
an inflation.
Attaining a sufficiently high reheat temperature after inflation
then requires relaxing the naturalness assumptions for
this field.
More appealing is the possibility that a flat direction
with renormalizable couplings to the visible sector
acts as the inflaton.
The reheat temperature is then naturally
low enough to avoid reproducing moduli and gravitinos from
thermal rescattering,
but above the weak scale so that baryogenesis can proceed through
weak scale processes or an Affleck-Dine mechansim.

A weak scale inflation can not naturally give rise to
cosmological density fluctuations of the magnitude required
for large scale structure formation.
However if the number of $e$-foldings during inflation
is sufficiently small, density fluctuations from an earlier
period of inflation can be preserved on large scales.
In fact it is perhaps more natural for the weak scale
inflation to persist for only a moderate number of $e$-foldings.
In conclusion, we have shown that same assumptions which
give the cosmological moduli problem also naturally
lead to its solution via inevitable weak scale inflation.
Within this solution successful baryogensis and the
preservation of large scale density fluctuations are also
possible.

{\bf Acknowledgements}:
We would like to thank T. Banks, M. Dine, A. Guth,
D. Kaplan,
and J. Primack for useful discussions, and the ITP for hospitality
during the Weak Interactions Workshop '94.

\refout
\end